\documentclass[journal,twoside,web]{ieeecolor}
\usepackage{generic}
\usepackage{cite}
\usepackage{amsmath,amssymb,amsfonts}
\usepackage{algorithmic}
\usepackage{graphicx}
\usepackage{algorithm,algorithmic}
\usepackage{hyperref}
\hypersetup{hidelinks=true}
\usepackage{textcomp}
\usepackage{float}
\def\BibTeX{{\rm B\kern-.05em{\sc i\kern-.025em b}\kern-.08em
    T\kern-.1667em\lower.7ex\hbox{E}\kern-.125emX}}
\graphicspath{{./figures/}}
\newtheorem{Definition}{Definition}     
\newtheorem{Theorem}{Theorem}               
\newtheorem{Remark}{Remark}                            
\newtheorem{Assumption}{Assumption}
\newtheorem{Lemma}{Lemma}

\begin{document}
\title{Explicit Solution of Tunable Input-to-State Safe-Based Controller Under High-Relative-Degree Constraints}
\author{Yan~Wei, Yu~Feng, Linlin~Ou, Yueying~Wang, Xinyi~Yu
\thanks{This work was supported in part by the National Natural Science Foundation of China (62203392), in part by the Natural Science Foundation of Zhejiang Province (LY23F030009), and in part by the the Baima Lake Laboratory Joint Funds of the Zhejiang Provincial Natural Science Foundation of China (LBMHD24F030002). (Corresponding author: )}
\thanks{Y. Wei, Y. Feng, L. Ou, and X.~Yu are with the College of Information Engineering, Zhejiang University of Technology, Hangzhou 310023, China (e-mail: weiyanok@zjut.edu.cn; yfeng@zjut.edu.cn; linlinou@zjut.edu.cn; yuxy@zjut.edu.cn).}
\thanks{Y. ~Wang is with the
School of Mechatronic Engineering and Automation, Shanghai University, Shanghai 200444, China (e-mail: yueyingwang@shu.edu.cn).}}
\maketitle

\begin{abstract}
This paper investigates the safety analysis and verification of nonlinear systems subject to high-relative-degree constraints and unknown disturbance. The closed-form solution of the high-order control barrier functions (HOCBF) optimization problem with and without a nominal controller is first provided, making it unnecessary to solve the quadratic program problem online and facilitating the analysis. Further, we introduce the concept of tunable input-to-state safety(ISSf), and a new tunable function in conjunction with HOCBF is provided. When combined with the existing ISSf theorem, produces controllers for constrained nonlinear systems with external disturbances. The theoretical results are proven and supported by numerical simulations.
\end{abstract}


\section{Introduction}
\label{sec:introduction}
\IEEEPARstart{T}{he} safety of system states is of critical
importance. System state safety is often formulated as an invariance problem related to a set of safe states. Numerous methods have been explored to address constraints, including barrier Lyapunov functions, and model predictive control techniques. However, the strong feasibility conditions and significant computational demands associated with these methods limit the practical applications. \par
The control barrier function (CBF) serves as an effective method for control design by establishing a straightforward condition that governs a desired safe set.  CBFs yield a state-dependent collection of inputs that ensure safety \cite{7782377}, which have been extensively studied and applied to various control systems, including robotic systems\cite{8913716}, autonomous vehicle systems\cite{SeoLee-192}. In \cite{XU201554}, a control Lyapunov function (CLF)-CBF-based quadratic program (QP) formulation was proposed. To date, the QP is still the main tool for implementing constrained control approaches. However, QP-based methods are not conducive to analysis, the calculation time cost is relatively long, and non-origin equilibrium points might be caused while the controller ensures system safety. In \cite{TanDimarogonas-445}, the design parameter that affects the equilibrium points was discussed. However, the effectiveness of the aforementioned CBF-based safety control approaches was typically demonstrated assuming that the systems model is accurate.\par
Ensuring the safety of systems in the presence of  uncertainties or disturbances poses significant challenges. The foundational work in \cite{9147463} proposed the notion of adaptive CBF, which model uncertainty was handled. Further, a robust CBF method was introduced to address disturbance \cite{Jankovic-415}. This may induce conservativeness and degrade the performance of a controller. Recent works to improve the robustness of CBFs using learning-based approaches such as Gaussian process regression\cite{9658123}, and neural networks\cite{9646174}. This depend on large amounts of quality data to improve robustness and generalization. Input-to-state safety (ISSf) serves as a valuable tool for analyzing systems affected by disturbances, having been initially defined in \cite{7799254}, and further elaborated in \cite{8003054}. Subsequently, input-to-state safe CBFs (ISSf-CBFs) were introduced, and a QP incorporating both CLFs and ISSf-CBFs was constructed, resulting in a unified safeguarding-stabilizing controller capable of operating under disturbances\cite{8405547}. In \cite{LyuXu-455}, a small-gain theorem was explored for the safety verification of interconnected systems. The studies mentioned above primarily focus on ISSf-CBFs with the relative degree of the system is one and does not apply to the general case of any relative degree.\par

Many constraints exhibit higher relative degrees in relation to the underlying system, such as position constraints in robotic systems. To address state constraints with arbitrary-order cases, a HOCBF was proposed, which was not restricted to exponential functions\cite{XiaoBelta-282}. In \cite{LyuXu-356}, a high relative-degree ISSf approach was introduced to quantify the influence of disturbance. Additionally, an ISSf-HOCBFs-based collision-free control strategy for surface vehicle was examined\cite{10234074}. However, these methods represent worst-case design approaches, leading to a conservative system behavior throughout the control process. To address this issue, a tunable ISSf-CBF was introduced in \cite{9448067}, allowing controllers to maintain safety guarantees in the presence of bounded  disturbances while mitigating conservatism. It is important to note that the aforementioned studies concentrate on cases. The tunable function may increase to arbitrarily large values, which could result in the controller approaching infinity.\par
In this article, we tackle the issue of affine systems with high relative degree constraints and disturbances. The main goal is to develop an ISSf theorem based on HOCBF for the safety verification of affine systems with disturbance. The contributions are summarized as follows:
\begin{enumerate}
\item The closed-form solution to the optimization problem for the HOCLF-HOCBF-based approach with or without a nominal controller is first provided, making it unnecessary to solve the QP problem online and enhancing the analysis.
\item The concept of a tunable ISSf-HOCBF is investigated, and a new tunable function is proposed to meet safety requirements in the presence of disturbances while reducing conservatism when integrated with the ISSf theorem.
\end{enumerate}
$\mathbf{Notations}$: $\mathcal{R}$ denotes the set of real numbers, $\mathcal{R}^n$ is the $n$-dimensional Euclidean space,  $\mathcal{R}_{\ge 0}$, $\mathcal{R}_{\le 0}$ are the set of nonnegative and nonpositive numbers, respectively. The Euclidean norm is denoted by $\|\ast\|$. $\mathcal{U}$
denotes a closed-control constraint set. $\mathcal{L}^m_\infty$ is a set of essentially bounded measurable functions. For any $\mu \in \mathcal{L}^m_\infty$, $\|\mu\|_k$ stands for the supremum norm of $\mu$ on an interval $k \subseteq \mathcal{R}_{\ge 0}$. The Lie derivatives of a function $B(x)$ for the system $\dot{x}=f(x)+g(x)u$ are denoted by $\mathcal{L}_fB=\bigtriangledown{B}^Tf(x) \in \mathcal{R}$.  $\mathcal{K}$ denotes class $\mathcal{K}$ functions, $\mathcal{K}^e$ denotes extended class $\mathcal{K}$ function, and $\mathcal{K}_\infty$ denotes infinity class $\mathcal{K}$ function(A continuous function $\alpha \in \mathcal{K}_\infty$ if $\alpha(0)=0$, $\alpha$ is strictly increasing, and $\lim_{q\to \infty} \alpha(q)=\infty$).
\section{Preliminaries}
\subsection{System Distribution}
Consider affine nonlinear systems in the form
\begin{equation} \label{System-Fun}
\begin{aligned}
\mathbf{\dot{x}}=\mathbf{f}(\mathbf{x})+\mathbf{g}(\mathbf{x})\mathbf{u},
\end{aligned}
\end{equation}
where $\mathbf{x}\in \mathcal{R}^n$ denotes the state, $\mathbf{f}: \mathcal{R}^n \to  \mathcal{R}^n$, and $\mathbf{g}: \mathcal{R}^n \to  \mathcal{R}^{n \times m}$ are assumed to be locally Lipschitz continuous on $\mathcal{R}^n$, $\mathbf{u}\in \mathcal{R}^m$ denotes control input.
\begin{Definition}[Forward Invariant\cite{7782377}]
A set  $\mathcal{C} \subset \mathcal{R}^n$ is forward invariant for system \eqref{System-Fun} if for every $\mathbf{x}(t_0)\in \mathcal{C}$,  $\mathbf{x}(t)\in \mathcal{C}$, for $\forall t >t_0$.
\end{Definition}
Supposed that there exists a set $\mathcal{C}$ makes the conditions to be forward invariant, which is defined as 
\begin{equation} \label{CBF-Set-1}
\mathcal{C}=\{\mathbf{x}\in \mathcal{R}^n: B(\mathbf{x}) \ge 0\},
\end{equation}
\begin{equation} \label{CBF-Set-2}
\partial{\mathcal{C}}=\{\mathbf{x}\in \mathcal{R}^n: B(\mathbf{x}) =0\},
\end{equation}
\begin{equation} \label{CBF-Set-3}
Int({\mathcal{C}})=\{\mathbf{x}\in \mathcal{R}^n: B(\mathbf{x}) >0\},
\end{equation}
where $B(\mathbf{x}): \mathcal{R}^n\to \mathcal{R}$ is a continuously differentiable function; $\partial{\mathcal{C}}$ and $Int({\mathcal{C}})$ are the boundary and interior, respectively. We call $C$ the safe set. It is said to be forward invariant and safe if for any initial condition $\mathbf{x}(t) \in \mathcal{C}, \forall t \ge 0$. We call the system \eqref{System-Fun} safe concerning the set $\mathcal{C}$.
\subsection{High Order Control Barrier Functions}
 Given a $r$-order differentiable function ${\mathbf{\varphi}}_{i}(\mathbf{x})$, a sequence of functions is defined as:\par
\begin{equation} \label{sequence-HOCBF}
\begin{aligned}%
{\mathbf{\varphi}}_{i}(\mathbf{x})=\dot{\mathcal{\varphi}}_{i-1}(\mathbf{x})+\beta_i(\mathcal{\varphi}_{i-1}(\mathbf{x})),~~i \in {1,...,r}
\end{aligned}
\end{equation}
where $\beta_i(\ast)\in \mathcal{K}^e$, ${\mathbf{\varphi}}_{0}(\mathbf{x})=\mathbf{B}(\mathbf{x})$.
Define the sets
\begin{equation} \label{HOCBF-set}
\begin{aligned}
S_{i-1}=\{\mathbf{x}\in \mathcal{R}^n: \mathcal{\varphi}_{i-1}(\mathbf{x}) \ge 0\}, i= 1\cdots,r.
\end{aligned}%
\end{equation}
We obtain that $\mathcal{S}_{0}=C$, and  $\mathcal{S}=\cap^r_{i=1}\mathcal{S}_{i-1}$ is the subsets of $\mathcal{C}$.
\begin{Definition}[HOCBF\cite{XiaoBelta-282}]
A function ${\mathbf{\varphi}}_{r}(\mathbf{x})$=: $\mathcal{R}^n \times [t_0, \infty) \to \mathcal{R}$ is a candidate HOCBF of relative degree $r$ for system if there exist differentiable smooth class $\mathcal{K}^e$ functions $\beta_i, i \in \{1,\dots,r\}$.
\begin{equation} \label{D-HOCBF}
\begin{aligned}
\sup_{u\in U}[\mathcal{L}_f&{{\mathcal{\varphi}}}_{r-1}(\mathbf{x})+\mathbf{\mathcal{L}_g}{{\mathcal{{\mathcal{{\mathcal{\varphi}}}}}}}_{r-1}(\mathbf{x})\mathbf{u}+\beta_r({\mathcal{{\mathcal{{\mathcal{\varphi}}}}}}_{r-1}(\mathbf{x}))]\ge 0
\end{aligned}
\end{equation}
for all $(\mathbf{x},t) \in \mathcal{S} = \mathcal{S}_0\cap, \cdots \cap \mathcal{S}_i\times [t_0, \infty)$, where $\mathcal{L}_f$ and $\mathcal{L}_g$ denote the partial Lie derivatives.
\end{Definition}\par
The above definition leads us to define the input sets: ${\mathbf{\varphi}}_{r}(\mathbf{x})$, $\forall \mathbf{x}\in \mathcal{S}$:
\begin{equation} \label{HOCBF-input-sets}
\begin{aligned}
	\mathcal{U}_{cbf}(\mathbf{x}):= &\{u\in\mathcal{R}^m|\mathcal{L}_f{\mathcal{\varphi}}_{r-1}(\mathbf{x})+\mathbf{\mathcal{L}_g}\mathcal{\varphi}_{r-1}(\mathbf{x})\mathbf{u}\\
	&+\beta_r(\mathcal{\varphi}_{r-1}(\mathbf{x}))\ge 0\}
\end{aligned}
\end{equation}
\begin{Lemma}[\cite{TanShaw-Cortez-296}]\label{Th-HOCBF-forward-invariant}
	Consider an HOCBF $B(\mathbf{x})$, ${\mathcal{\varphi}}_{i-1}(\mathbf{x})$, $1\le i \le r$ defined in \eqref{sequence-HOCBF}. If $\mathbf{x}(t_0)\in \mathcal{S}(t_0)$, then any locally Lipschitz continuous controller $\mathbf{u}: \mathcal{R}^n\to \mathcal{R}^m$ such that $u\in \mathcal{U}_{cbf}$, $\forall t \ge t_0$ will render the set $\mathcal{S}(t)$ forward invariant for the system \eqref{System-Fun}.
\end{Lemma}
\subsection{High Order Control Lyapunov Functions}
Supposed that there exists a set $\mathcal{D} \in \mathcal{R}^n$ makes the conditions to be forward invariant, which is defined as follows:
\begin{equation} \label{CLF-Set-1}
	\mathcal{D}=\{\mathbf{x}\in \mathcal{R}^n: \dot{V}(\mathbf{x}) \le 0\},
\end{equation}
\begin{equation} \label{CLF-Set-2}
	\partial{\mathcal{D}}=\{\mathbf{x}\in \mathcal{R}^n: \dot{V}(\mathbf{x}) =0\},
\end{equation}
\begin{equation} \label{CLF-Set-3}
	Int({\mathcal{D}})=\{\mathbf{x}\in \mathcal{R}^n: \dot{V}(\mathbf{x}) <0\},
\end{equation}
where $V(\mathbf{x}): \mathcal{R}^n\to \mathcal{R}$ is a continuously differentiable Lyapunov function. We call $D$ the stable set. It is said to be forward invariant and stable if for any initial condition $\mathbf{x}(0) \in \mathcal{D}$, $\mathbf{x}(t) \in \mathcal{D}, \forall t \ge 0$. We call the system \eqref{System-Fun} stable concerning the set $\mathcal{D}$.\par
Similar to HOCBF, we define a sequence of functions:
\begin{equation} \label{sequence-HOCLF}
	\begin{aligned}
	    \phi_{0}(\mathbf{x})&=-\dot{V}(\mathbf{x})-\eta({V}(\mathbf{x})),\\
		{\mathbf{\phi}}_{i}(\mathbf{x})&=\dot{\mathbf{\mathbf{\phi}}}_{i-1}(\mathbf{x})+\alpha_i(\phi_{i-1}(\mathbf{x})),~i \in {1,...,r}
	\end{aligned}
\end{equation}
where ${V}(\mathbf{x})$ is a Lyapunov function, $\eta(\ast) \in \mathcal{K}$, and $\alpha_i(\ast) \in \mathcal{K}^e$.
Define the sets
\begin{equation} \label{HOCLF-set}
	\begin{aligned}
		\mathcal{Z}_{i}=\{\mathbf{x}\in \mathcal{R}^n: \mathcal{\phi}_{i-1}(\mathbf{x}) \ge 0\}, i\in 1,\cdots,r
	\end{aligned}%
\end{equation}
we obtain that $\mathcal{Z}_{0}=D$, and  $\mathcal{Z}=\cap^r_{i=1}\mathcal{Z}_{i-1}$ is the subsets of $\mathcal{D}$.
\begin{Definition}[HOCLF\cite{WOS:001196709200041}]\label{DD-HOCLF}
	Let functions ${\mathbf{\phi}}_{i}(\mathbf{x})$ and sets $\mathcal{Z}_{i}$ be defined by \eqref{sequence-HOCLF} and \eqref{HOCLF-set}, respectively. 
	$V(0) = 0$, $V (\mathbf{x})> 0$ $\forall x \in \mathcal{R}^n, x\neq 0$ is a HOCLF of relative degree $r$ for the system if there exist differentiable class $\mathcal{K}^e$ functions $\alpha_i, i \in \{1,\dots,r\}$ such that
	\begin{equation} \label{D-HOCLF}
		\begin{aligned}
			\sup_{u\in U}[\mathcal{L}_f&{\mathbf{\phi}}_{r-1}(\mathbf{x})+\mathbf{\mathcal{L}_g}{\mathbf{\mathbf{\phi}}}_{r-1}(\mathbf{x})\mathbf{u}+\alpha_r(\phi_{r-1}(\mathbf{x}))] \ge 0.
		\end{aligned}
	\end{equation}
\end{Definition}
For an HOCLF, we define the input set:
\begin{equation} \label{Control-HOCLF}
	\begin{aligned}
		\mathcal{U}_{clf}(\mathbf{x}):= &\{\mathbf{u}\in\mathcal{R}^m| \mathcal{L}_f{\mathbf{\mathbf{\phi}}}_{r-1}(\mathbf{x})+\mathbf{\mathcal{L}_g}{\mathbf{\mathbf{\phi}}}_{r-1}(\mathbf{x})\mathbf{u}\\
		&+\alpha_r(\phi_{r-1}(\mathbf{x}))\ge 0\}.
	\end{aligned}%
\end{equation}
\begin{Lemma}\label{FI-Th-HOCLF}
Consider an HOCLF $V(x)$, ${\mathbf{\phi}}_{i-1}(\mathbf{x})$, $1 \le i \le r$ defined in \eqref{sequence-HOCLF} with the associated sets $\mathcal{Z}_{i}, i\in \{1,\cdots,r\}$. Then, any locally Lipschitz continuous controller $u \in \mathcal{U}_{clf}$ will render the set $\mathcal{Z}$ forward invariant for the system \eqref{System-Fun}, then the set is asymptotically stable.
\end{Lemma}
\begin{proof}
See in the Appendix \ref{Appendix-HOCLF}.
\end{proof}
\section{Explicit solution of tunable ISSf HOCBF}
\subsection{Tunable Input to State Safe HOCBF}
Consider affine nonlinear systems with external disturbance:
\begin{equation} \label{System-Fun-d}
\begin{aligned}
\mathbf{\dot{x}}=\mathbf{f}(\mathbf{x})+\mathbf{g}_1(\mathbf{x})\mathbf{u}+\mathbf{g}_2(\mathbf{x})\mathbf{d}(t),
\end{aligned}
\end{equation}
where $\mathbf{g}_1, \mathbf{g}_2: \mathcal{R}^n \to  \mathcal{R}^{n \times m}$ are assumed to be locally Lipschitz continuous on $\mathcal{R}^n$, $\mathbf{d}: \mathcal{R}^n \to \mathcal{R}^{m}$ denotes time-varying bounded matched/unmatched disturbance. The objective is to design a safety controller that ensures safety in the presence of disturbances. We have the following assumption for external disturbance:\par
\begin{Assumption}
The external disturbance $\mathbf{d}$ is bounded and piecewise continuous in time. That is, the following inequation holds $\forall t\in [0, \infty)$:
\begin{equation}
\|\mathbf{d}\|_\infty= {\sup}_{t\ge 0} \|\mathbf{d}(t)\| \le \gamma <\infty,
\end{equation}
\end{Assumption}
where $\gamma>0$ is a constant. Note that the set $\mathcal{C}$ is safe if it is forward invariant. We say that $\mathcal{C}$ is ISSf if a slightly larger set $ \mathcal{C}_{\gamma} \supseteq \mathcal{C}$ is forward invariant with disturbance.
Consider the set $\mathcal{C}_{\gamma} \subset \mathcal{R}$ defined as
\begin{equation} \label{ISSF-Set-1}
	\mathcal{C}_{\gamma}=\{\mathbf{x}\in \mathcal{R}^n: B(\mathbf{x}) +\varrho(B(\mathbf{x}), \gamma)\ge 0\},
\end{equation}
\begin{equation} \label{ISSF-Set-2}
	\partial{\mathcal{C}}_{\gamma}=\{\mathbf{x}\in \mathcal{R}^n: B(\mathbf{x}) +\varrho(B(\mathbf{x}), \gamma)=0\},
\end{equation}
\begin{equation} \label{ISSF-Set-3}
	Int({\mathcal{C}_{\gamma}})=\{\mathbf{x}\in \mathcal{R}^n: B(\mathbf{x}) +\varrho(B(\mathbf{x}), \gamma)>0\},
\end{equation}
with $\varrho: \mathcal{R} \times \mathcal{R}_{\ge 0} \to \mathcal{R}_{\ge 0}$ satisfying  $\varrho(a,\ast) \in \mathcal{K}_\infty$ for all $a \in \mathcal{R}$. $\partial{\mathcal{C}}_{\gamma}$ and $Int({\mathcal{C}_{\gamma}})$ are the boundary and interior of $\mathcal{C}$, respectively.  With this construction in mind, we have the definition of TISSf.
\begin{Definition}(Tunable input-to-state safety)
	Given a set $\mathcal{C} \subset \mathcal{R}^n$ defined by \eqref{CBF-Set-1}--\eqref{CBF-Set-3} for a continuously differentiable function $B(x): \mathcal{R}^n \to \mathcal{R}$.  The system \eqref{System-Fun-d} is tunable input-to-state safe (TISSf) with respect to the
	set $\mathcal{C}$, if there exists $\varrho: \mathcal{R} \times \mathcal{R}_{\ge 0} \to \mathcal{R}_{\ge 0}$ satisfying $\varrho(a,\ast)\in \mathcal{K}_\infty$ for all  $a\in \mathcal{R}$ and $\varrho(\ast,b)$ continuously differentiable for all $b\in \mathcal{R}_\infty$ such that for all $\gamma\ge 0$ and $d$ satisfying $\|d\|_\infty\le \gamma$, the set $\mathcal{C}_{\gamma}$ in \eqref{ISSF-Set-1}--\eqref{ISSF-Set-3} is forward invariant.
	If the system \eqref{System-Fun-d} is TISSf concerning the
	set $\mathcal{C}$, the set $\mathcal{C}$ is referred to as a TISSf set.
\end{Definition}
Define the sets
\begin{equation} \label{ISSf-HOCBF-set}
	\begin{aligned}
		\mathcal{Q}_{i-1}=\{\mathbf{x}\in \mathcal{R}^n: \mathcal{\varphi}_{i-1}(\mathbf{x})+\varrho(\mathcal{\varphi}_{i-1}(\mathbf{x}), \gamma) \ge 0\},
	\end{aligned}%
\end{equation}
we obtain that $\mathcal{Q}_{0}=\mathcal{C}_\gamma$, and  $\mathcal{Q}=\cap^r_{i=1}\mathcal{Q}_{i-1}$ is the subsets of $\mathcal{C}_\gamma$.\par
For system \eqref{System-Fun-d}, TISSf-HOBF is defined as follows:
\begin{Definition}[TISSf-HOBF] 
Given a set $\mathcal{Q} \subset \mathcal{R}^n$, the function $\mathcal{\varphi}_{i-1}(\mathbf{x})$ is called an TISSf-HOBF that is $r$-th order differentiable for system \eqref{System-Fun-d} on  $\mathcal{C}$ if there exists $\gamma>0$, $\varrho\in \mathcal{K}_\infty$ such that \eqref{sequence-HOCBF} and 
\begin{equation} \label{ISSF-HOBF}
\mathcal{\varphi}_{r}(\mathbf{x})\ge -\varrho(\mathcal{\varphi}_{r-1}(\mathbf{x}), \gamma),
\end{equation}
for all $\mathbf{x} \in \mathcal{R}^n$ and satisfying $\|d\|_{\infty}\le \gamma$.
\end{Definition}
\begin{Lemma}[\cite{LyuXu-356}]
Let $\mathcal{Q}$ be the 0-super level set of a continuously
differentiable $\mathcal{\varphi}_{i-1}(\mathbf{x})$. If $\mathcal{\varphi}_{i-1}(\mathbf{x})$ is an TISSf-HOBF for system \eqref{System-Fun-d} on $\mathcal{Q}$, the set $\mathcal{Q}$ is forward invariant, and the system is TISSf on the set $\mathcal{C}$.
\end{Lemma}\par
The right-hand side of \eqref{ISSF-HOBF}, $\varrho(\mathcal{\varphi}_{r-1}(\mathbf{x}), \gamma)$ denotes the tunable function, which guarantees the establishment of the new set $\mathcal{C}_{\gamma}$, this further ensures the establishment of TISSf set $\mathcal{C}$. For an affine control system, TISSf-HOCBF is defined as:
\begin{Definition}[TISSf-HOCBF]
Let ${\mathcal{\varphi}}_{r}(\mathbf{x})$ be defined by \eqref{sequence-HOCBF}. Given a set $\mathcal{Q} \subset \mathcal{R}^n$, the function $\mathcal{\varphi}_{r-1}(\mathbf{x})$ is an TISSf-HOCBF of relative degree $r$ for system \eqref{System-Fun-d} on $\mathcal{Q}$, if there exists $\gamma>0$, $ \beta_r \in \mathcal{K}^e_\infty$ such that
\begin{equation}\label{ISSf-HOCBF}
\begin{aligned}
\sup_{u\in \mathcal{R}^m}&[\mathcal{L}_f\mathcal{\varphi}_{r-1}(\mathbf{x})+\mathbf{\mathcal{L}_{g1}}\mathcal{\varphi}_{r-1}(\mathbf{x})\mathbf{u}]\\
&\ge-\beta_r(\mathcal{\varphi}_{r-1}(\mathbf{x}))+ \frac{\|\mathbf{\mathcal{L}_{g2}}{\mathcal{\varphi}}_{r-1}(\mathbf{x})\|^2_2}{\epsilon(\mathcal{\varphi}_{r-1}(\mathbf{x}))},
\end{aligned}
\end{equation}
where $\epsilon: \mathcal{R} \to \mathcal{R}_{> 0}$ is a continuously differentiable function.
\end{Definition}\par
The set of all control values exists that satisfy \eqref{ISSf-HOCBF}:
\begin{equation} \label{Control-ISSf-HOCBF}
\begin{aligned}
\mathcal{U}_{TISSf}=&\big\{\mathbf{u}\in \mathcal{R}^m|\mathcal{L}_f{{\mathcal{{\mathcal{{\mathcal{\varphi}}}}}}}_{r-1}(\mathbf{x})+\mathbf{\mathcal{L}_{g1}}{\mathcal{\varphi}}_{r-1}(\mathbf{x})\mathbf{u}\\
&\ge-\beta_r({\mathcal{{\mathcal{{\mathcal{\varphi}}}}}}_{r-1}(\mathbf{x})) +\frac{\|\mathbf{\mathcal{L}_{g2}}{\mathcal{\varphi}_{r-1}(\mathbf{x})\|^2_2}}{\epsilon(\mathcal{\varphi}_{r-1}(\mathbf{x}))}\big\}.
\end{aligned}%
\end{equation}
$\epsilon(\mathcal{\varphi}_{r-1}(\mathbf{x}))$ is used to tune the balance between performance and safety.
In this section, we concentrate on the situation where $B(\mathbf{x})$ is high-relative-degree. The external disturbance explicitly appears until $B(\mathbf{x})$ is differentiated $r > 1$ times. Then, the point-wise solution of HOCBF-QP-based method with or without disturbance is investigated.
\subsection{Explicit Solution of HOCBF-Based Controller Without Disturbance}
The minimum-norm HOCLF-HOCBF-based controller is given by the following quadratic program:
\begin{equation} \label{CLF-HOCBF}
	\begin{aligned}
		&\tau_{QP}(\mathbf{x})=\arg\min\limits_{\mathbf{u}\in\mathcal{R}^m}\frac{1}{2}\|\mathbf{u}\|^2_2+\frac{1}{2}\rho\sigma^2\\
		&s.t.~\mathcal{L}_f{\mathbf{\mathbf{\phi}}}_{r-1}(\mathbf{x})+\mathbf{\mathcal{L}_g}{\mathbf{\mathbf{\phi}}}_{r-1}(\mathbf{x})\mathbf{u}+\alpha_r(\mathcal{\phi}_{r-1}(\mathbf{x})) \ge \sigma,\\
		&s.t.~\mathcal{L}_f{\mathcal{\varphi}}_{r-1}(\mathbf{x})+\mathbf{\mathcal{L}_g}{\mathcal{\varphi}}_{r-1}(\mathbf{x})\mathbf{u}+\beta_r({\mathcal{{\mathcal{{\mathcal{\varphi}}}}}}_{r-1}(\mathbf{x}))\ge 0,
	\end{aligned}
\end{equation}
where $\sigma$ is a slack variable used to relax the HOCLF constraint, $\rho$ is a positive constant. The first constraint is a stabilization constraint relaxed with a slack variable $\sigma$, the second one is a safety constraint. Then, the quadratic program in \eqref{CLF-HOCBF} is always feasible.\par

Inspired by \cite{TanDimarogonas-445}, the closed-form solution of \eqref{CLF-HOCBF} is given as the following Theorem:
\begin{Theorem}\label{Th-CLF-HOCBF-1}
	Denote the control input given as a solution of \eqref{CLF-HOCBF} as $\mathbf{\tau}_{QP}(\mathbf{x})$. Then, the solution to the quadratic program in \eqref{CLF-HOCBF} is given by
\begin{equation} \label{CLF-HOCBF-Control-2}
\begin{aligned}
			\mathbf{\tau}_{QP}(\mathbf{x})=\left\{\begin{array}{lcr}
				\mathbf{0},                         ~~~~~~~~~~~~~~~~~~~~~~~~~~~\mathbf{x}\in \Omega_1 \cup  \Omega_2\\
				-\frac{\Gamma_b}{\omega_1}\mathbf{\mathcal{L}_g}\mathcal{\varphi}^T_{r-1},    ~~~~~~~~~~~~~~~\mathbf{x}\in \Omega_3 \\
				-\frac{\Gamma_v}{1/\rho+\omega_2}\mathbf{\mathcal{L}_g}{\mathbf{\mathbf{\phi}}}^T_{r-1},   ~~~~~~~~~~\mathbf{x}\in \Omega_4 \cup \Omega_5\\
				-q_1\mathbf{\mathcal{L}_g}{\mathbf{\mathbf{\phi}}}^T_{r-1}+q_2\mathbf{\mathcal{L}_g}{{\mathcal{{\mathcal{{\mathcal{\varphi}}}}}}}^T_{r-1},                          ~\mathbf{x}\in \Omega_6\\		
			\end{array}\right.
\end{aligned}%
\end{equation}
where $\Gamma_v:=\mathcal{L}_f{\mathbf{\mathbf{\phi}}}_{r-1}+\alpha_r(\phi_{r-1})$, $\Gamma_b:=\mathcal{L}_f{{\mathcal{{\mathcal{{\mathcal{\varphi}}}}}}}_{r-1}+\beta_r({\mathcal{\varphi}}_{r-1})$, $\omega_1={\mathbf{\mathcal{L}_g}{\mathcal{\varphi}}_{r-1}\mathbf{\mathcal{L}_g}{{\mathcal{\varphi}}^T_{r-1}}}$, $\omega_2={\mathbf{\mathcal{L}_g}{\mathcal{\phi}}_{r-1}\mathbf{\mathcal{L}_g}{{\mathcal{\phi}}^T_{r-1}}}$, $\omega_3=\mathbf{\mathcal{L}_g}{\mathbf{\mathbf{\phi}}}_{r-1}\mathbf{\mathcal{L}_g}{{\mathcal{{\mathcal{{\mathcal{\varphi}}}}}}}^T_{r-1}$,\\ 
$	\left[\begin{array}{c}
	q_1 \\
	q_2
\end{array}\right]=	\left[\begin{array}{cc}
	1/\rho+\omega_2 & -\omega_3 \\
	-\omega_3 & \omega_1
\end{array}\right]^{-1}\left[\begin{array}{c}
	\Gamma_v \\
	-\Gamma_b
\end{array}\right]$. The set of $\Omega_i$ are given as
\begin{equation} \label{Omega_1}
	\begin{aligned}
		&\Omega_1=\{\mathbf{x}\in \mathcal{R}^n: \Gamma_v>0, \Gamma_b>0\},
	\end{aligned}
\end{equation}
\begin{equation} \label{Omega_2}
	\begin{aligned}
		&\Omega_2=\{\mathbf{x}\in \mathcal{R}^n: \Gamma_v>0, \Gamma_b=0,~ \mathbf{\mathcal{L}_g}\mathcal{\varphi}_{r-1}=0 \} ,
	\end{aligned}
\end{equation}
\begin{equation} \label{Omega_3}
	\begin{aligned}
		\Omega_3=&\{\mathbf{x}\in \mathcal{R}^n: \Gamma_b\le0, \Gamma_v\omega_1-\Gamma_b\omega_3>0\},
	\end{aligned}
\end{equation}
\begin{equation} \label{Omega_4}
	\begin{aligned}
		\Omega_4=&\{\mathbf{x}\in \mathcal{R}^n: \Gamma_v \le 0, \Gamma_v\mathbf{\mathcal{L}_g}{{\mathcal{{\mathcal{{\mathcal{\varphi}}}}}}}\mathbf{\mathcal{L}_g}{\mathbf{\mathbf{\phi}}}^T_{r-1}\\
		&-\Gamma_b(1/\rho+\omega_1)<0\},
	\end{aligned}%
\end{equation}
\begin{equation} \label{Omega_5}
	\begin{aligned}
		&\Omega_5=\{\mathbf{x}\in \mathcal{R}^n: \Gamma_v \le 0, \Gamma_b=0, \mathbf{\mathcal{L}_g}{\mathbf{\mathbf{\varphi}}}_{r-1}=0\},
	\end{aligned}
\end{equation}
\begin{equation} \label{Omega_6}
	\begin{aligned}
		\Omega_6=\{\mathbf{x}\in \mathcal{R}^n: \Gamma_v\omega_3-\Gamma_b(1/\rho+\omega_2)\ge0,~ \mathbf{\mathcal{L}_g}{\mathbf{\mathbf{\varphi}}}_{r-1} \neq 0\},
	\end{aligned}
\end{equation}
\end{Theorem}
\begin{proof}  
See in the Appendix \ref{Appendix-A}.
\end{proof} 
\begin{Remark}
	The expression in \eqref{CLF-HOCBF} guarantees that a closed-form solution can be derived for HOCBF-QP-based control problems. Theorem \ref{Th-CLF-HOCBF-1} offers a general form solution for CBF-QP-based control problems applicable to any relative degree.
\end{Remark}\par
Given an HOCBF ${{\mathcal{{\mathcal{{\mathcal{\varphi}}}}}}}_{r-1}$, the set of control inputs with nominal controller that can be obtained:
\begin{equation} \label{HOCBF-Control-1}
	\begin{aligned}
		&\mathbf{\tau}_{QP}(\mathbf{x})=\arg\min\limits_{\mathbf{u}\in\mathcal{R}^m}\frac{1}{2}\| \textbf{u}-\mathbf{\tau}_n(\mathbf{x})\|^2_2\\
		&s.t.~\mathcal{L}_f{\mathcal{\varphi}}_{r-1}(\mathbf{x})+\mathbf{\mathcal{L}_g}{\mathcal{\varphi}}_{r-1}(\mathbf{x})\mathbf{u}+\beta_r({\mathcal{{\mathcal{{\mathcal{\varphi}}}}}}_{r-1}(\mathbf{x}))\ge 0,
	\end{aligned}
\end{equation}
where  the nominal controller  $\mathbf{\tau}_n(\mathbf{x}):\mathcal{R}^n\to\mathcal{R}^m$ Lipschitz continuous in $\mathbf{x}$. The nominal controller is only modified when it does not satisfy safety requirements.\par
The following theorem provides a closed-form solution to the
optimization problem in \eqref{HOCBF-Control-1}.
\begin{Theorem}\label{Th-HOCBF-1}
	Suppose that we have a continuous controller $\mathbf{\tau}_n: \mathcal{R}^n \to \mathcal{R}^m$, denotes the nominal controller, that does not necessarily ensure the closed-loop system \eqref{System-Fun} without disturbance is safe with respect to the set $\mathcal{C}$, but achieves a desired degree of performance.
	The optimization based safety controller $\tau_n: \mathcal{R}^n \to \mathcal{R}^m$ is defined as.
	\begin{equation} \label{HOCBF-sol}
		\begin{aligned}
			\mathbf{\tau}_{QP}(x)=\tau_{n}(\mathbf{x})+{\xi}(\mathbf{x})\mathbf{\mathcal{L}_g}{{\mathcal{{\mathcal{{\mathcal{\varphi}}}}}}}^T_{r-1},
		\end{aligned}
	\end{equation}
	where the function ${\xi}(\mathbf{x}) : \mathcal{R}^n \to \mathcal{R}$ is defined as
\begin{equation} \label{HOCBF-Control-2}
\begin{aligned}
\xi(\mathbf{x})=\left\{\begin{array}{lcr}
0,~~~~~\mathbf{\mathcal{L}_g}\mathcal{\varphi}_{r-1}\tau_{n}+\Gamma_b(\mathbf{x}) \ge 0 \cup \mathcal{L}_g\mathcal{\varphi}_{r-1}=0\\
-\frac{\Gamma_b+\mathbf{\mathcal{L}_g}{\mathcal{\varphi}}^T_{r-1}\mathbf{\tau}_n}{\|\mathbf{\mathcal{L}_g}{\mathcal{\varphi}}^T_{r-1}\|^2_2},~~~~~~~~~~ otherwise
\end{array}\right.
\end{aligned}
\end{equation}
\end{Theorem}
\begin{proof} 
See in the Appendix \ref{Appendix-B}.
\end{proof}
\subsection{Tunable ISSf HOCBF-Based Controller}
TISSf-HOCBF is defined as a way for safety-critical control analysis.
\begin{Theorem}\label{Th-TISSf-HOCBF-ForwardInvariant}
	If $B(\mathbf{x})$, ${{\mathcal{{\mathcal{{\mathcal{\varphi}}}}}}}_{i-1}(\mathbf{x})$ is a  TISSf-HOCBF for \eqref{System-Fun-d}  on $\mathcal{Q}$  with continuously differentiable function $\epsilon: \mathcal{R}\to  \mathcal{R}_{>0}$, $\beta_r\in \mathcal{K}^e_\infty$ such that,  $\beta^{-1}_r\in \mathcal{K}^e_\infty$ and $\epsilon$ satisfies:
	\begin{equation} \label{TISSf-HOCBF-Control}
		\begin{aligned}
			\frac{d\epsilon }{d r}(\epsilon({\mathcal{\varphi}}_{r-1}(\mathbf{x})))\ge0,
		\end{aligned}
	\end{equation}
	then for any Lipschitz continuous controller with $\mathbf{u}$ and  for all disturbance $\mathbf{d}$ satisfying
	$\|\mathbf{d}\|_\infty \le \gamma$. Then, the system \eqref{System-Fun-d} is safe with respect to $\mathcal{Q}$ with $\varrho \in \mathcal{K}_{\infty}$ defined as:
	\begin{equation} \label{TISSf-HOCBF-varrho}
		\begin{aligned}
			\varrho({\mathcal{\varphi}}_{r-1}(\mathbf{x})), \gamma)=-\beta^{-1}_{r}\bigg(-\frac{\epsilon({\mathcal{\varphi}}_{r-1}(\mathbf{x}))\gamma^2}{4}\bigg).
		\end{aligned}
	\end{equation}
This implies $\mathcal{Q}$ is an TISSf set.
\end{Theorem}
\begin{proof}  
See in the Appendix \ref{Appendix-C}.
\end{proof} 
The function $\epsilon({\mathcal{\varphi}}_{r-1}(\mathbf{x}))$ is proposed as follows:
\begin{equation} \label{epsilon-B}
\begin{aligned}
\epsilon({\mathcal{\varphi}}_{r-1}(\mathbf{x}))=\frac{1}{\epsilon_{0}+e^{{\varsigma}{\mathcal{\varphi}}_{r-1}(\mathbf{x})}}
\end{aligned}
\end{equation}
with parameters $\epsilon_{0}>0$, $\varsigma\ge 0$.
\begin{Remark}
The tunable function proposed is different from the function in \cite{AlanTaylor-313,8405547}. The value of $\epsilon ({\mathcal{\varphi}}_{r-1}(\mathbf{x}))$ is limited to the range of $\epsilon ({\mathcal{\varphi}}_{r-1}(\mathbf{x}))\in (0,\frac{1}{\epsilon_{0}})$ to ensure that the control input is not too large.
\end{Remark}
The minimum-norm TISSf-HOCLF-HOCBF-based controller is given by the following quadratic program:
\begin{equation} \label{ISSf-CLF-HOCBF}
\begin{aligned}
&\bar{\tau}_{QP}(\mathbf{x})=\arg\min\limits_{\mathbf{u}\in\mathcal{R}^m}\frac{1}{2}\|\mathbf{u}\|^2_2+\frac{1}{2}\rho\sigma^2\\
s.t.&~\mathcal{L}_f{\mathbf{\mathbf{\phi}}}_{r-1}(\mathbf{x})+\mathbf{\mathcal{L}_{g1}}{\mathbf{\mathbf{\phi}}}_{r-1}(\mathbf{x})\mathbf{u}+\alpha_r(\phi(\mathbf{u})) \le \sigma,\\
s.t.&~\mathcal{L}_f{{\mathcal{{\mathcal{{\mathcal{\varphi}}}}}}}_{r-1}(\mathbf{x})+\mathbf{\mathcal{L}_{g1}}{{\mathcal{{\mathcal{{\mathcal{\varphi}}}}}}}_{r-1}(\mathbf{x})\mathbf{u}+\beta_r({\mathcal\varphi}_{r-1}(\mathbf{x}))\\
&\ge \frac{\|\mathbf{\mathcal{L}_{g2}}\mathcal{\varphi}_{r-1}(\mathbf{x})\|^2_2}{\epsilon({\mathcal{\varphi}}_{r-1}(\mathbf{x}))},
\end{aligned}
\end{equation}
where $\sigma$ is a slack variable used to relax the CLF constraint, $\rho$ is a positive constant, then the quadratic program in \eqref{CLF-HOCBF} is always feasible.
\begin{Theorem}\label{Th-CLF-HOCBF-Distubance}
	Denote the control input given as a solution of \eqref{ISSf-CLF-HOCBF} as $\mathbf{\bar{\tau}}_{QP}(\mathbf{x})$. Then, the solution to the quadratic program in \eqref{ISSf-CLF-HOCBF} is given by
\begin{equation} \label{CLF-HOCBF-sol-D}
		\begin{aligned}
			\bar{\tau}_{QP}(x)=\left\{\begin{array}{lcr}
				0,                         ~~~~~~~~~~~~~~~~~~~~~~~~~~~~~~\mathbf{x}\in \Omega_{d1} \cup  \Omega_{d2}\\
				-\frac{\Gamma_b}{\omega_1}\mathbf{\mathcal{L}_{g1}}{{\mathcal{{\mathcal{{\mathcal{\varphi}}}}}}}^T_{r-1},                               ~~~~~~~~~~~~~~~~\mathbf{x}\in \Omega_{d3} \\
				-\frac{\Gamma_v}{1/\rho+\omega_2}\mathbf{\mathcal{L}_{g1}}{\mathbf{\mathbf{\phi}}}^T_{r-1},                                ~~~~~~~~~~~\mathbf{x}\in \Omega_{d4} \cup \Omega_{d5}\\
				-q_1\mathbf{\mathcal{L}_{g1}}{\mathbf{\mathbf{\phi}}}^T_{r-1}+q_2\mathbf{\mathcal{L}_{g1}}{{\mathcal{{\mathcal{{\mathcal{\varphi}}}}}}}^T_{r-1},                                                                     ~\mathbf{x}\in \Omega_{d6}\\		
			\end{array}\right.
		\end{aligned}
	\end{equation}
where $\Gamma_v(\mathbf{x}):=\mathcal{L}_f{\mathbf{\mathbf{\phi}}}_{r-1}(\mathbf{x})+\alpha_r(\phi(\mathbf{x}))$, $\Gamma_d(\mathbf{x}):=\Gamma_b(\mathbf{x})+\varrho({\mathcal{\varphi}}_{r-1}(\mathbf{x})), \gamma)$,\\ $\left[\begin{array}{c}
	q_3 \\
	q_4
\end{array}\right]=	\left[\begin{array}{cc}
	1/\rho+\omega_2 & -\omega_3 \\
	-\omega_3 & \omega_1
\end{array}\right]^{-1}\left[\begin{array}{c}
	\Gamma_v \\
	\Gamma_d
\end{array}\right]$. The set of $\Omega_{di}$ are given as
\begin{equation} \label{Omega_d1}
	\begin{aligned}
		&\Omega_{d1}=\{\mathbf{x}\in \mathcal{R}^n: \Gamma_v>0, \Gamma_d>0\} ,
	\end{aligned}
\end{equation}
\begin{equation} \label{Omega_d2}
	\begin{aligned}
		&\Omega_{d2}=\{\mathbf{x}\in \mathcal{R}^n: \Gamma_v>0, \Gamma_d=0, \mathbf{\mathcal{L}_{g1}}{{\mathcal{{\mathcal{{\mathcal{\varphi}}}}}}}_{r-1}=0 \} ,
	\end{aligned}
\end{equation}
\begin{equation} \label{Omega_d3}
	\begin{aligned}
		\Omega_{d3}=&\{\mathbf{x}\in \mathcal{R}^n: \Gamma_d\le0, \Gamma_v\omega_1-\Gamma_b\omega_3>0\},
	\end{aligned}
\end{equation}
\begin{equation} \label{Omega_d4}
	\begin{aligned}
	\Omega_{d4}=\{\mathbf{x}\in \mathcal{R}^n: \Gamma_v \ge 0, \Gamma_v\omega_1-\Gamma_d(1/\rho+\omega_1)<0\},
	\end{aligned}%
\end{equation}
\begin{equation} \label{Omega_d5}
	\begin{aligned}
		&\Omega_{d5}=\{\mathbf{x}\in \mathcal{R}^n: \Gamma_v \le 0, \Gamma_d=0, \mathbf{\mathcal{L}_{g1}}{\mathbf{\mathbf{\varphi}}}_{r-1}=0\},
	\end{aligned}
\end{equation}
\begin{equation} \label{Omega_d6}
\begin{aligned}
\Omega_{d6}=&\{\mathbf{x}\in \mathcal{R}^n: \Gamma_v\mathbf{\mathcal{L}_{g1}}{\mathbf{\mathbf{\phi}}}_{r-1}\mathbf{\mathcal{L}_{g1}}\mathcal{\varphi}^T_{r-1}-\Gamma_d(1/\rho\\
&+\omega_2\ge0, \mathbf{\mathcal{L}_{g1}}\mathbf{\varphi}_{r-1} \neq 0\},
\end{aligned}
\end{equation}
\end{Theorem}
\begin{proof}
It is similar to the proof in Theorem \ref{Th-CLF-HOCBF-1}.
\end{proof}
Given an HOCBF $\mathcal{\varphi}_{r-1}$, the set of control inputs that can be obtained: 
\begin{equation} \label{ISSF-HOCBF-Control}
\begin{aligned}
\bar{\tau}_{QP}&(\mathbf{x})=\arg\min\limits_{u\in\mathcal{R}^m}\frac{1}{2}\| \mathbf{u}-\tau_n(\mathbf{x})\|^2_2\\
s.t.&~\mathcal{L}_f{{\mathcal{{\mathcal{{\mathcal{\varphi}}}}}}}_{r-1}(\mathbf{x})+\mathbf{\mathcal{L}_{g1}}{{\mathcal{{\mathcal{{\mathcal{\varphi}}}}}}}_{r-1}(\mathbf{x})\mathbf{u}+\beta_r({\mathcal\varphi}_{r-1}(\mathbf{x}))\\
&\ge \frac{\|\mathbf{\mathcal{L}_{g2}}\mathcal{\varphi}_{r-1}(\mathbf{x})\|^2_2}{\epsilon({\mathcal{\varphi}}_{r-1}(\mathbf{x}))},
\end{aligned}
\end{equation}
where $\mathbf{\tau}_n(\mathbf{x}):\mathcal{R}^n \to\mathcal{R}^m$ Lipschitz continuous in $\mathbf{x}$.\par
The following theorem provides a closed-form solution to the optimization problem defining this controller.
\begin{Theorem}\label{Th-HOCBF-Disturbance}
Suppose we have a continuous controller $\mathbf{\tau}_n(x): \mathcal{R}^n \to \mathcal{R}^m$, which denotes the nominal controller, that does not necessarily ensure the closed-loop system \eqref{System-Fun-d} with disturbance is safe with respect to the set $\mathcal{C}$. Then, the optimization problem in \eqref{ISSf-HOCBF} is feasible for any $\mathbf{x} \in \mathcal{R}^n$
and has a closed-form solution given by
\begin{equation} \label{ISSf-HOCBF-sol-1}
\begin{aligned}
\bar{\tau}_{QP}(\mathbf{x})=\tau_{n}(\mathbf{x})+\bar{\xi}(\mathbf{x})\mathbf{\mathcal{L}_{g1}}{{\mathcal{{\mathcal{{\mathcal{\varphi}}}}}}}^T_{r-1},
\end{aligned}
\end{equation}
where the function $\bar{\xi}(x): \mathcal{R}^n \to \mathcal{R}$ is defined as
\begin{equation} \label{ISSf-HOCBF-Control-2}
\begin{aligned}
\bar{\xi}(\mathbf{x})=	\left\{\begin{array}{lcr}
0,~~~\mathbf{\mathcal{L}_{g1}}\mathcal{\varphi}_{r-1}\tau_{n}+\Gamma_d(\mathbf{x}) \ge 0 \cup \mathcal{L}_{g1}\mathcal{\varphi}_{r-1}=0\\
-\frac{\Gamma_d+\mathbf{\mathcal{L}_{g1}}{\mathcal{\varphi}}^T_{r-1}\tau_n}{\|\mathbf{\mathcal{L}_{g1}}{\mathcal{\varphi}}^T_{r-1}\|^2_2},~~~~~~~~ otherwise				
\end{array}\right.
\end{aligned}
\end{equation}
$\bar{\xi}(\mathbf{x})$ is continuous and $\bar{\xi}(\mathbf{x})\in U_{TISSf} (\mathbf{x})$, $\forall \mathbf{x}\in \mathcal{R}^n$.
\end{Theorem}
The proof is similar to the proof of Theorem \ref{Th-HOCBF-1}.
\section{Simulation Studies}
A simulation example of the proposed control method is presented to verify its effectiveness.
\begin{equation} \label{pendulum-spring-cart system}
	\begin{aligned}
		\left\{\begin{array}{lcr}
			\dot{x}_{i,1}={x}_{i,2}+d_{i,1}(t)\\
			\dot{x}_{i,2}=\frac{g}{{\omega}l}{x}_{i,1}-\frac{m}{M}\sin{x_{i,1}}-\frac{r_1\vartheta(r_1-{\omega}l)}{{\omega}l^2}(x_{i,1}-x_{3-i,1})\\
			+\frac{r_2\vartheta(r_1-{\omega}l)}{{\omega}l^2}+\frac{1}{{\omega}l^2}\tau_i+\frac{1}{{\omega}l^2}d_{i,2}(t),     (i=1,~2),\\
		\end{array}\right.
	\end{aligned}%
\end{equation}
where $x_{i,1}=\theta_i$, $x_{i,2}=\dot{\theta}_i$ are angular and angular velocity, respectively, $\tau_i$ represents the control torque, $M$ denotes the mass of the car, $d_i(t)=[d_{i,1}(t);d_{i,2}(t)]$ is the external disturbance which is satisfied $|d_i(t)|\le \bar{d}$, ${\omega}=m/(M+m)$, $\vartheta$ denotes the spring constant, $r_1$ represents the distance from the pivot of the spring to the bottom of the pendulum, $r_2$ is the distance between the cars, $g$ is the gravitational acceleration. The value of system parameters: $g=9.8 m/s^2$, $l=1m$,  $\vartheta=1$, $M=15 kg$, $m=5 kg$, $r_1=0.75m$, $r_2=2m$. The desired signal: $x_{d}=[\sin(t); -\sin(t+\pi/4)]$. Denote $\underline{\theta}_i$ as the lower bound of $x_{i,1}$. The parameters of tunable function: $\epsilon_{0}=0.06$, $\varsigma = 200$. Unless otherwise specified, the value of the parameters in the following two cases are the same. The control objective is to drive $x_{i,1}$ to the desired angular $x_{d}$ while avoiding an obstacle and robust with disturbance.\par

$\mathbf{Case 1}$: Explicit solution of TISSf-HOCLF-HOCBF.\par
The closed-form solution of HOCLF-HOCBF in Theorem \ref{Th-CLF-HOCBF-Distubance} is verified in this case. Denote $\mathbf{z}_{1}=\mathbf{x}_{1}-\mathbf{x}_{d}$, the CLF candidate is defined as $V(x)=\frac{1}{2}\mathbf{z}^2_{1}$.  Denote $\mathbf{e}_{1}=\mathbf{x}_{1}-\mathbf{\underline{\theta}}$, We define $B(x) =\mathbf{e}_{1}$ as an HOCBF with $r = 2$. $\underline{\theta}=[-0.3; -0.3]$. Matched disturbance is considered, which is given as $d_2=[-10; -10]$. Two CBF-based methods are provided to demonstrate the effectiveness of the proposed TISSf-HOCLF-HOCBF controller. Simulation comparison 1: robust CBF-based approch mentioned in \cite{Jankovic-415}(Robust HOCBF); Simulation comparison 2: the HOCBF-based controller without a tunable function(HOCBF without TISSf). \par
\begin{figure}[!t]
	\begin{center}
		\includegraphics[width=2.8in]{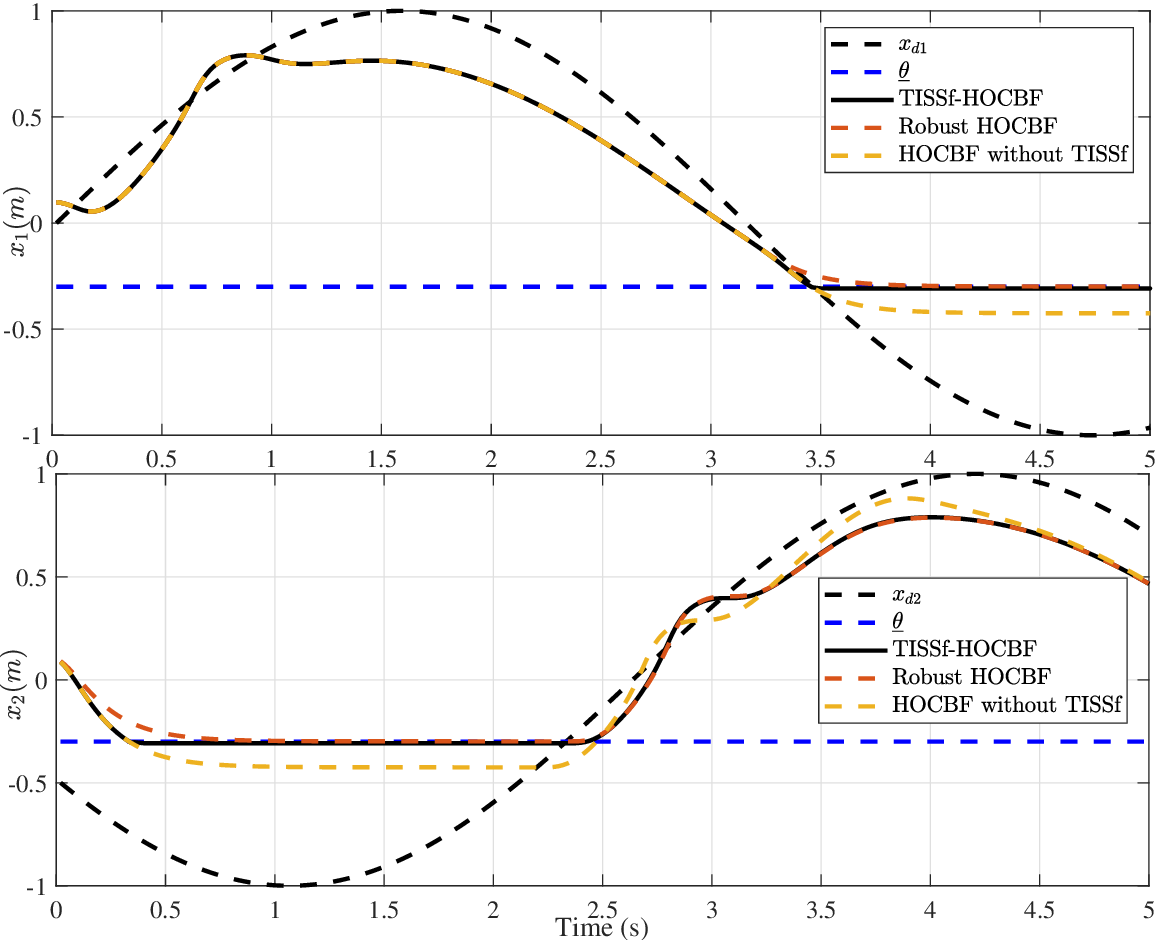}     
		\caption{Trajectories comparison of $x_1$ under constraints (Case 1).}  
		\label{Case1x}                                 
	\end{center}                                       
\end{figure}
\begin{figure}[!t]
	\begin{center}
		\includegraphics[width=2.8in]{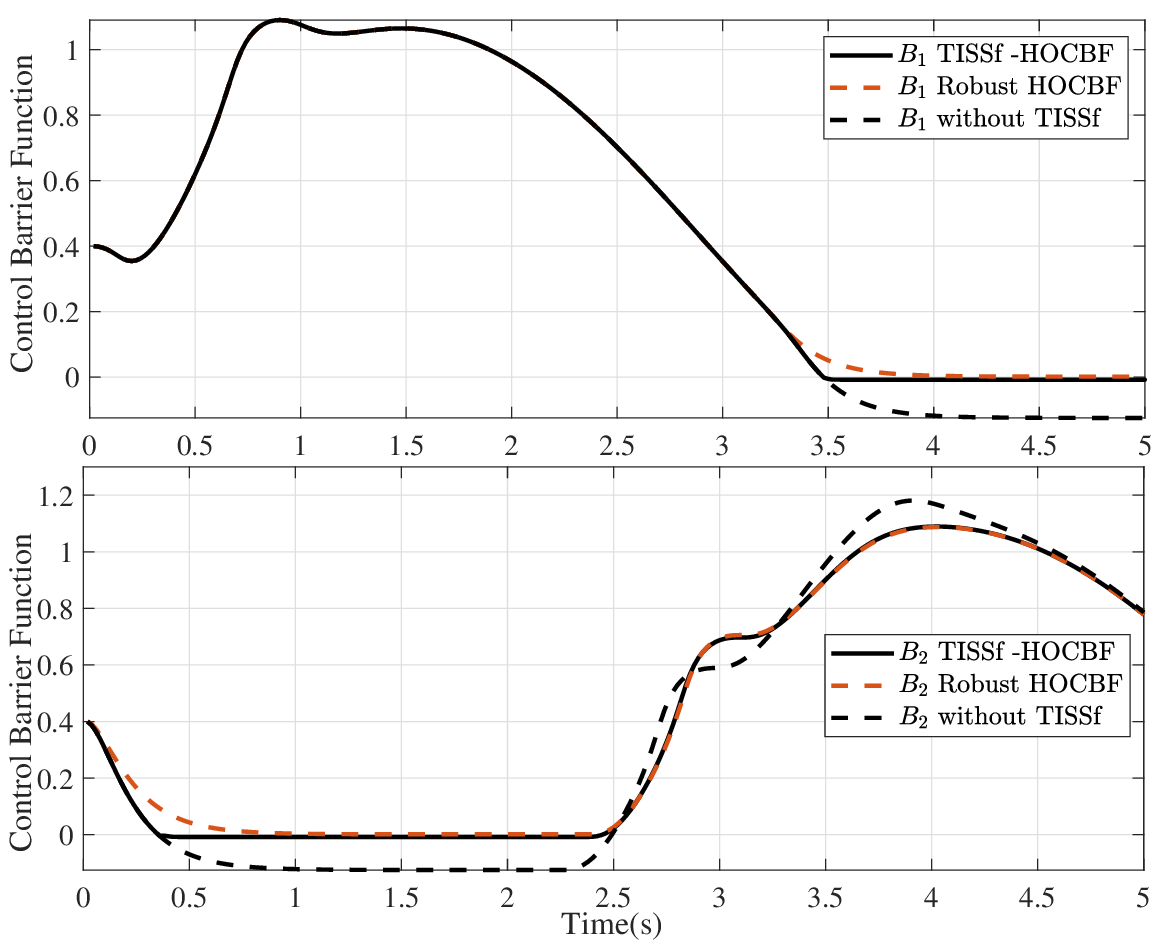}         
		\caption{Trajectories of CBF comparison (Case 1). }                    
		\label{Case1hISSfCLFCBF}                                    
	\end{center}                                                    
\end{figure}

The simulation results are presented in Figs. \ref{Case1x}--\ref{Case1hISSfCLFCBF}. Fig. \ref{Case1x} illustrates the comparison of trajectory tracking performance under constraints and disturbance, while Fig. \ref{Case1hISSfCLFCBF} displays the trajectories for the comparison of CBFs. As indicated by Fig. \ref{Case1x},  the close solution obtained in Theorem \ref{Th-CLF-HOCBF-1} successfully achieves both trajectory tracking and safety control. \par


$\mathbf{Case 2}$: TISSf-HOCBF with nominal controller.\par
The closed-form solution of TISSf-HOCBF
in Theorem \ref{Th-HOCBF-Disturbance} is verified in this case. $\underline{\theta}=[-0.25; 0.25]$. We design a nominal tracking controller with the classical backstepping technique. Define the tracking error function as 
\begin{equation} \label{nominal-controller-error}
	\begin{aligned}
		z_{i,1}=x_{i,1}-x_{i,d},~ z_{i,2}=x_{i,2}-\upsilon_{i,1},
	\end{aligned}
\end{equation}
where $\upsilon_{i,1}$ denotes the control input. Then the nominal controller is designed as 
\begin{equation} \label{nominal-VirtualController}
	\begin{aligned}
		\upsilon_{i,1}=-k_1z_{i,1}+x_{i,r},
	\end{aligned}
\end{equation}
\begin{equation} \label{nominal-FnalController}
	\begin{aligned}
		u^{nom}_{i}=-f_{i,2}+{\omega}l^2(-k_2z_{i,2}+z_{i,1}+\dot{\upsilon}_{i,1}),
	\end{aligned}
\end{equation}
where $k_1>0$, $k_2>0$ are two parameters to be determined. The stability analysis of the nominal controller is omitted. In this subsection, we choose $k_1=k_2=10$  Unmatched disturbance is considered, which is given as $d_1=[0.5; 0.5]$. Two CBF-based methods are provided to demonstrate the effectiveness of the proposed controller. Simulation comparison 1: robust CBF-based approch with classical backstepping technique(Robust HOCBF BSC); Simulation comparison 2: the HOCBF-based controller without a tunable function(HOCBF BSC without TISSf).\par
\begin{figure}[!t]
	\begin{center}
		\includegraphics[width=2.8in]{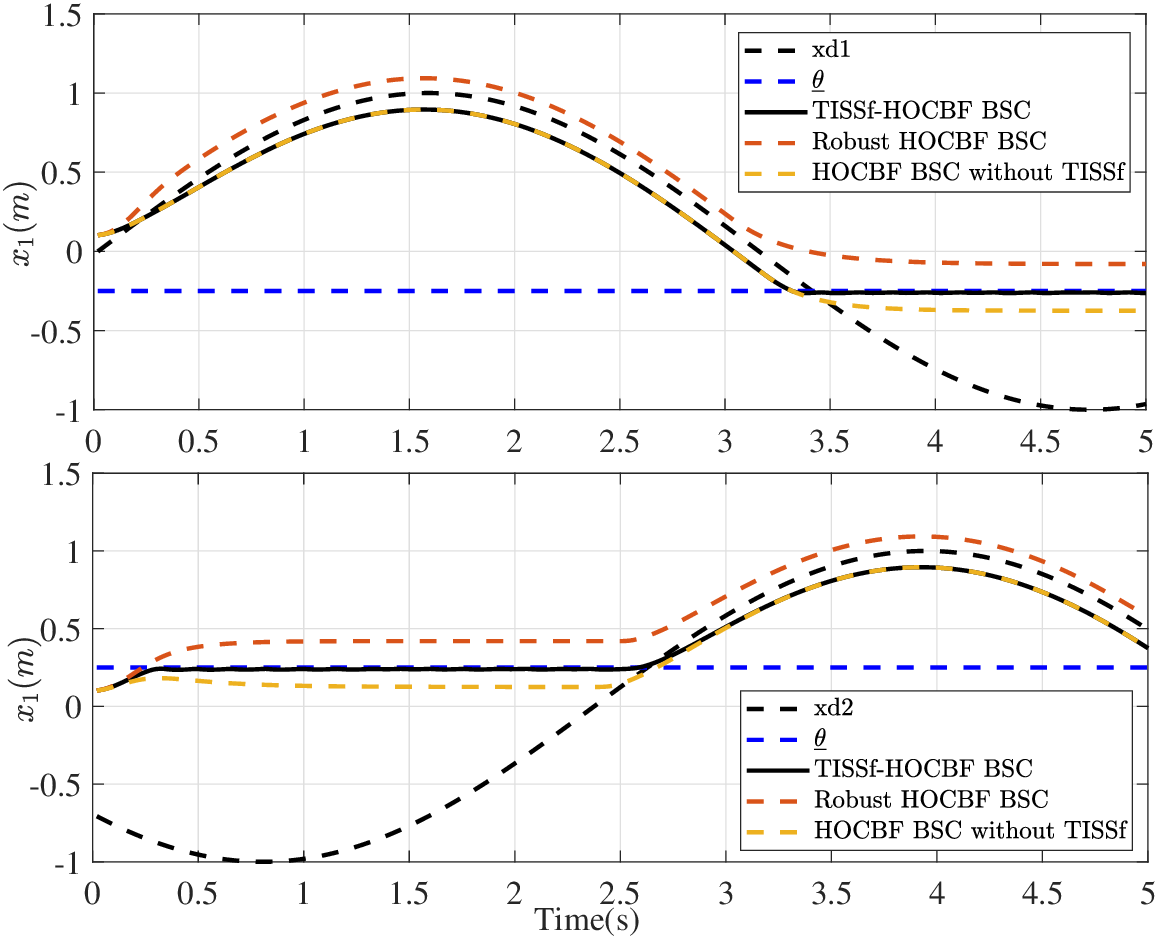}    
		\caption{Trajectories comparison of $x_1$ under constraints (Case 2).}  
		\label{Case2x1-x}                                 
	\end{center}                                 
\end{figure}
\begin{figure}[!t]
\begin{center}
\includegraphics[width=2.8in]{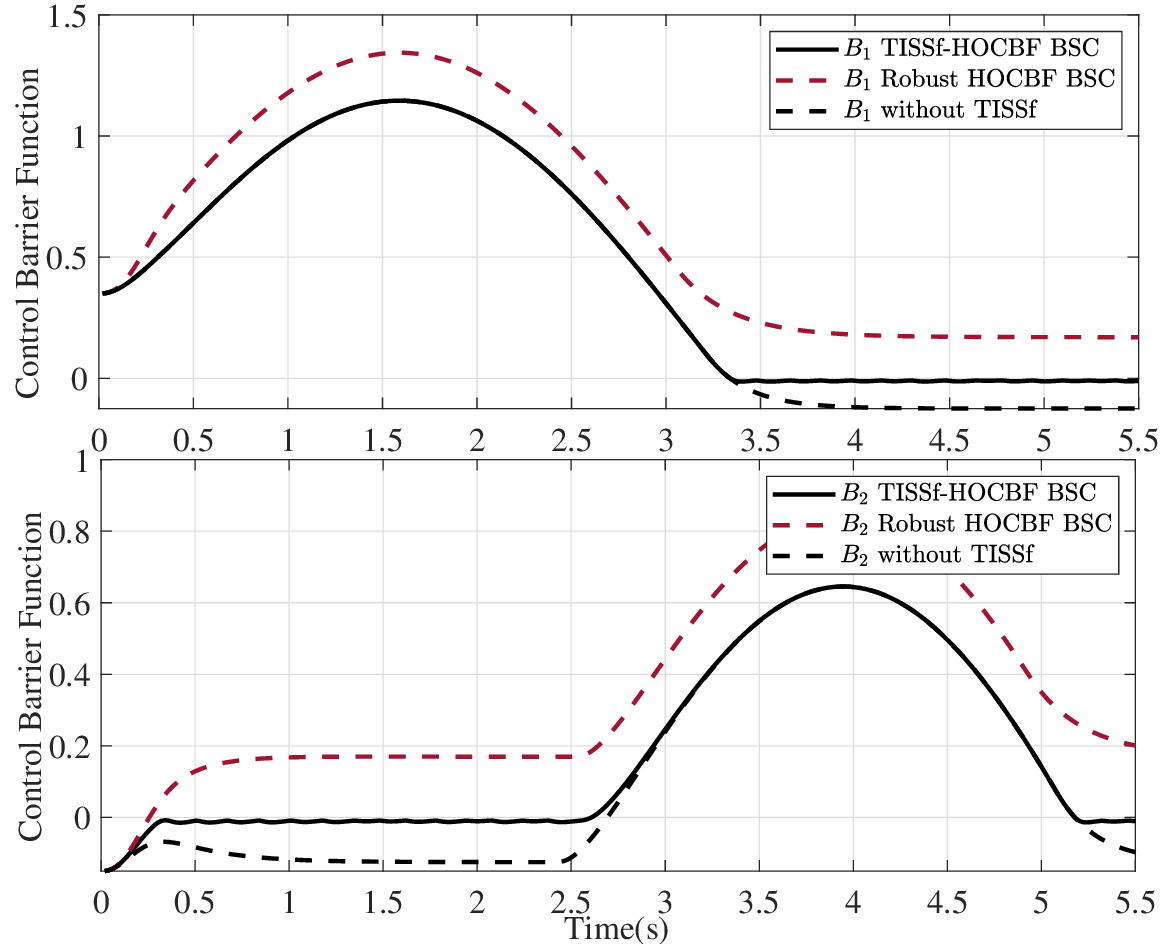}
\caption{Trajectories of CBF comparison (Case 2).}
\label{Case2-h}                                 
\end{center}                                 
\end{figure}
The simulation results are presented in Figs. \ref{Case2x1-x}--\ref{Case2-h}.  Fig. \ref{Case2x1-x} illustrates the trajectory tracking performance under constraints and disturbances. Fig. \ref{Case2-h} depicts the trajectories for the comparison of CBFs. From Fig. \ref{Case2x1-x}, it is evident that the closed-loop solution derived in Theorem \ref{Th-HOCBF-Disturbance} can successfully achieves both trajectory tracking and safety control.\par
From the above two cases, we can see that the designed method can guarantee the safety of the system under both matched and unmatched disturbances. In comparison to the tracking performance by robust HOCBF or the traditional method without tunable function, the proposed TISSf-HOCBF effectively ensures safety while reducing conservatism.
\section{Conclusion}
In this work, a TISSf theorem has been developed within the framework of HOCBF for the safety verification of nonlinear systems. We have presented a closed-form solution to the HOCBF optimization problem, applicable both with and without a nominal controller. Additionally, the introduction of ISSf-HOCBF offers a tunable control approach for constrained nonlinear systems subject to external disturbances. The theoretical results was verified by two case simulations. Future research will aim to extend the ISSf theorem to discrete systems.

\appendices
\section{Proof of Lemma \ref{FI-Th-HOCLF}} \label{Appendix-HOCLF}
If $V (\mathbf{x})$ is an HOCLF, then we have $\mathcal{\phi}_{r}(\mathbf{x})\ge 0, \forall t \in [t_0, \infty)$. Form Definition \ref{DD-HOCLF} and the equation in \eqref{sequence-HOCLF}, we can obtain the following inequation:
 \begin{equation} \label{ineq-HOCLF}
 	\begin{aligned}
 		\dot{\mathbf{\mathbf{\phi}}}_{r-1}(\mathbf{x})+\alpha_r(\phi_{r-1}(\mathbf{x}))\ge 0.
 	\end{aligned}%
 \end{equation}
 The solution of \eqref{ineq-HOCLF} is a class $\mathcal{KL}$ function $\hbar$ such that the following inequation holds:
  \begin{equation} \label{ineq-HOCLF-2}
 	\begin{aligned}
 		\dot{\mathbf{\mathbf{\phi}}}_{r-1}(\mathbf{x})\ge\hbar(\phi_{r-1}(\mathbf{x(t_0)},t_0)).
 	\end{aligned}
 \end{equation}
Since we start in the converging set, we have $x(t_0)\in \mathcal{Z}(t_0) \in \mathcal{Z}_{r}(t_0)$. Then, we have
\begin{equation} \label{ineq-HOCLF-3}
\begin{aligned}
\dot{\mathbf{\phi}}_{r-1}&(\mathbf{x(t_0)},t_0)+\alpha_r(\phi_{r-1}(\mathbf{x(t_0)},t_0))\\
=&\alpha_r(\phi_{r-1}(\mathbf{x(t_0)},t_0))\ge 0.
\end{aligned}
\end{equation}
Then, we have
  \begin{equation} \label{ineq-HOCLF-4}
	\begin{aligned}
		\dot{\mathbf{\mathbf{\phi}}}_{r-1}(\mathbf{x})\ge 0.
	\end{aligned}
\end{equation}
Iteratively, we can get $x(t) \in \mathcal{Z}_{i}(t),  \forall i\in[1, r],  \forall t\in[0, \infty]$. Therefore, the set $\mathcal{Z}:= \cap_{i=1}^r \mathcal{Z}_i$ is forward invariant for the system \eqref{System-Fun}.\par
Since the set $\mathcal{Z}$ is forward invariant, we have 
 \begin{equation} \label{ineq-HOCLF-5}
\mathcal{\phi}_{0}(\mathbf{x}) =-\dot{V}(\mathbf{x})-\eta({V}(\mathbf{x}))\ge 0, \forall t>t_0.
\end{equation}
Then, because $\dot{V}(\mathbf{x})$ is a Lyapunov function, the origin of system\eqref{System-Fun} is asymptotically stable.
\section{Proof of Theorem \ref{Th-CLF-HOCBF-1}} \label{Appendix-A}    
The Lagrangian associated to the QP \eqref{CLF-HOCBF} is $\ell_{1}=\frac{1}{2}\|\mathbf{u}\|^2_2+\frac{1}{2}\rho\sigma^2+\mu_1(\Gamma_v+\mathbf{\mathcal{L}_g}{{\mathcal{{\mathcal{{\mathcal{\varphi}}}}}}}_{r-1}\mathbf{u}-\sigma)-\mu_2(\Gamma_b+\mathbf{\mathcal{L}_g}{\mathbf{\mathbf{\phi}}}_{r-1}\mathbf{u})$, where $\mu_1$, $\mu_2$ are the Lagrangian multipliers. The Karush–Kuhn–Tucker (KKT) conditions are
\begin{equation} \label{Appendix-CLF-HOCBF-1}
	\begin{aligned}%
		\frac{\partial{\ell_{1}}}{\partial{\mathbf{u}}}=\mathbf{u}+\mu_1\mathbf{\mathbf{\mathcal{L}_g}{{\mathcal{{\mathcal{{\mathcal{\varphi}}}}}}}^T_{r-1}}-\mu_2\mathbf{\mathbf{\mathcal{L}_g}{\mathbf{\phi}}^T_{r-1}}=0,
	\end{aligned}
\end{equation}
\begin{equation} \label{Appendix-CLF-HOCBF-2}
\begin{aligned}%
\frac{\partial{\ell_{1}}}{\partial{\sigma}}=\rho\sigma-\mu_1=0,
\end{aligned}
\end{equation}
\begin{equation} \label{Appendix-CLF-HOCBF-3}
	\begin{aligned}%
		\mu_1(\Gamma_v+\mathbf{\mathbf{\mathcal{L}_g}}{\mathbf{\mathbf{\phi}}}^T_{r-1}\mathbf{u}-\sigma)=0,
	\end{aligned}
\end{equation}
\begin{equation} \label{Appendix-CLF-HOCBF-4}
	\begin{aligned}%
		\mu_2(\Gamma_b+\mathbf{\mathcal{L}_g}{{\mathcal{{\mathcal{{\mathcal{\varphi}}}}}}}_{r-1}\mathbf{u})=0.
	\end{aligned}
\end{equation}
\begin{equation} \label{Appendix-CLF-HOCBF-5}
\begin{aligned}%
\mu_1\ge0,~\mu_2\ge0.
\end{aligned}
\end{equation}
\begin{equation} \label{Appendix-CLF-HOCBF-6}
\begin{aligned}%
\mathcal{L}_f{\mathbf{\mathbf{\phi}}}_{r-1}(\mathbf{x})+\mathbf{\mathcal{L}_g}{\mathbf{\mathbf{\phi}}}_{r-1}(\mathbf{x})\mathbf{u}+\alpha_r(\mathcal{\phi}_{r-1}(\mathbf{x})) \ge \sigma,\\
\end{aligned}
\end{equation}
\begin{equation} \label{Appendix-CLF-HOCBF-7}
\begin{aligned}%
\mathcal{L}_f{\mathcal{\varphi}}_{r-1}(\mathbf{x})+\mathbf{\mathcal{L}_g}{\mathcal{\varphi}}_{r-1}(\mathbf{x})\mathbf{u}+\beta_r({\mathcal{{\mathcal{{\mathcal{\varphi}}}}}}_{r-1}(\mathbf{x}))\ge 0,
\end{aligned}
\end{equation}

$\mathbf{Case 1}$: The case of $\mathbf{x}\in \Omega_1$. In this case,
we have
\begin{equation} \label{Appendix-CLF-HOCBF-case1-1}
	\begin{aligned}%
		\Gamma_v+\mathbf{\mathcal{L}_g}{{\mathcal{{\mathcal{{\mathcal{\varphi}}}}}}}^T_{r-1}\mathbf{u}>\sigma,
	\end{aligned}
\end{equation}
\begin{equation} \label{Appendix-CLF-HOCBF-case1-2}
	\begin{aligned}%
		\Gamma_b+\mathbf{\mathcal{L}_g}{\mathbf{\mathbf{\phi}}}^T_{r-1}\mathbf{u}>0,
	\end{aligned}
\end{equation}
\begin{equation} \label{Appendix-CLF-HOCBF-case1-3}
	\begin{aligned}%
		\mu_1=0,~\mu_2=0.
	\end{aligned}
\end{equation}
From this case, we have $\mathbf{u}^{\ast}=\mathbf{0}$.\par
$\mathbf{Case 2}$: The CLF constraint is inactive and the HOCBF constraint is active. In this case, we have
\begin{equation} \label{Appendix-CLF-HOCBF-case2-1}
	\begin{aligned}%
		\Gamma_v+\mathbf{\mathcal{L}_g}{\mathbf{\mathbf{\phi}}}^T_{r-1}\mathbf{u}>\sigma,
	\end{aligned}
\end{equation}
\begin{equation} \label{Appendix-CLF-HOCBF-case2-2}
	\begin{aligned}%
		\Gamma_b+\mathbf{\mathcal{L}_g}{{\mathcal{{\mathcal{{\mathcal{\varphi}}}}}}}^T_{r-1}\mathbf{u}=0,
	\end{aligned}
\end{equation}
\begin{equation} \label{Appendix-CLF-HOCBF-case2-3}
	\begin{aligned}%
		\mu_1=0,~\mu_2\ge0.
	\end{aligned}
\end{equation}
From \eqref{Appendix-CLF-HOCBF-2}, we have $\sigma=\mu_1/\rho=0$. When $\mathbf{\mathcal{L}_g}{{\mathcal{{\mathcal{{\mathcal{\varphi}}}}}}}_{r-1}=0$ , from \eqref{Appendix-CLF-HOCBF-1}, \eqref{Appendix-CLF-HOCBF-case2-3} we have $\mathbf{u}^{\ast}=\mathbf{0}$. In this case, $\mathbf{x} \in  \Omega_3$.  When $\mathbf{\mathcal{L}_g}{{\mathcal{{\mathcal{{\mathcal{\varphi}}}}}}}_{r-1}\neq0$, from \eqref{Appendix-CLF-HOCBF-1}, \eqref{Appendix-CLF-HOCBF-case2-2}, we have $\mathbf{u}^{\ast}=-\frac{\Gamma_v}{\omega_1}\mathbf{\mathcal{L}_g}{{\mathcal{{\mathcal{{\mathcal{\varphi}}}}}}}^T_{r-1}$. In this case $x\in  \Omega_3$. \par
$\mathbf{Case 3}$:  The CLF constraint is active and the HOCBF constraint is inactive. In this case, we have
\begin{equation} \label{Appendix-CLF-HOCBF-case3-1}
	\begin{aligned}%
		\Gamma_v+\mathbf{\mathcal{L}_g}{\mathbf{\mathbf{\phi}}}^T_{r-1}\mathbf{u}=\sigma,
	\end{aligned}
\end{equation}
\begin{equation} \label{Appendix-CLF-HOCBF-case3-2}
	\begin{aligned}%
		\Gamma_b+\mathbf{\mathcal{L}_g}{{\mathcal{{\mathcal{{\mathcal{\varphi}}}}}}}^T_{r-1}\mathbf{u}=0,
	\end{aligned}
\end{equation}
\begin{equation} \label{Appendix-CLF-HOCBF-case3-3}
	\begin{aligned}%
		\mu_1\ge0,~\mu_2=0.
	\end{aligned}
\end{equation}
From \eqref{Appendix-CLF-HOCBF-1}, \eqref{Appendix-CLF-HOCBF-case3-3}, we have $\mathbf{u}=-\mu_1\mathbf{\mathcal{L}_g}{{\mathcal{{\mathcal{{\mathcal{\varphi}}}}}}}^T_{r-1}$, thus we have
\begin{equation}\label{Appendix-CLF-HOCBF-case3-4}
	\mathbf{\mathcal{L}_g}{{\mathcal{{\mathcal{{\mathcal{\varphi}}}}}}}_{r-1}\mathbf{u}=-\mathbf{\mathcal{L}_g}{{\mathcal{{\mathcal{{\mathcal{\varphi}}}}}}}_{r-1}\mu_1\mathbf{\mathcal{L}_g}{{\mathcal{{\mathcal{{\mathcal{\varphi}}}}}}}^T_{r-1}.
\end{equation}
Substituting \eqref{Appendix-CLF-HOCBF-case3-4} into \eqref{Appendix-CLF-HOCBF-case3-1}, yields
\begin{equation}\label{Appendix-CLF-HOCBF-case3-5}
	\mu_1=\frac{\Gamma_v}{1/\rho+\omega_1},
\end{equation}
\begin{equation}\label{Appendix-CLF-HOCBF-case3-6}
	\mathbf{u}=-\frac{\Gamma_v}{\omega_1}\mathbf{\mathcal{L}_g}{{\mathcal{{\mathcal{{\mathcal{\varphi}}}}}}}^T_{r-1}.
\end{equation}
In this case $\mathbf{x}\in  \Omega_4$.\par
$\mathbf{Case 4}$:  The CLF constraint and the HOCBF constraint are active. In this case, we have
\begin{equation} \label{Appendix-CLF-HOCBF-case4-1}
	\begin{aligned}%
		\Gamma_v+\mathbf{\mathcal{L}_g}{\mathbf{\mathbf{\phi}}}^T_{r-1}\mathbf{u}=\sigma,
	\end{aligned}
\end{equation}
\begin{equation} \label{Appendix-CLF-HOCBF-case4-2}
	\begin{aligned}%
		\Gamma_b+\mathbf{\mathcal{L}_g}{{\mathcal{{\mathcal{{\mathcal{\varphi}}}}}}}^T_{r-1}\mathbf{u}=0,
	\end{aligned}
\end{equation}
\begin{equation} \label{Appendix-CLF-HOCBF-case4-3}
	\begin{aligned}%
		\mu_1\ge0,~\mu_2\ge0.
	\end{aligned}
\end{equation}
From \eqref{Appendix-CLF-HOCBF-1}, \eqref{Appendix-CLF-HOCBF-2}, one has $\mathbf{u}=-\mu_1\mathbf{\mathcal{L}_g}{\mathbf{\mathbf{\phi}}}^T_{r-1}+\mu_2\mathbf{\mathcal{L}_g}{{\mathcal{{\mathcal{{\mathcal{\varphi}}}}}}}^T_{r-1}$,  $\sigma=\mu_1/\rho$. Substituting $\mathbf{u}$, $\rho$ into \eqref{Appendix-CLF-HOCBF-case4-1},
\eqref{Appendix-CLF-HOCBF-case4-1}, we have
\begin{equation} \label{Appendix-CLF-HOCBF-case4-4}
	\left[\begin{array}{cc}
		\frac{1}{\rho}+\omega_2 & -\omega_3 \\
		-\omega_3 & \omega_1
	\end{array}\right]\left[\begin{array}{c}
		\mu_1 \\
		\mu_2\end{array}\right]=\left[\begin{array}{c}
		\Gamma_v \\
		-\Gamma_b
	\end{array}\right].
\end{equation}
Denote $\Delta:=\det\bigg(\left[\begin{array}{cc}
	\frac{1}{\rho}+\omega_2  & -\omega_3 \\
	-\omega_3 & \omega_1
\end{array}\right]\bigg)$, we know that $\Delta= 0$ if and only if $\mathbf{\mathcal{L}_g}{{\mathcal{{\mathcal{{\mathcal{\varphi}}}}}}}_{r-1}$ for any $\rho>0$. 
When $\mathbf{\mathcal{L}_g}\mathcal{\varphi}_{r-1}=0$, we have
\begin{equation} \label{Appendix-CLF-HOCBF-case4-5}
	\mu_1=\frac{\Gamma_v}{\omega_1}.
\end{equation}
$\mu_2$ could be any positive scalar, and $\Gamma_b=0$. In view of \eqref{Appendix-CLF-HOCBF-1}, we have 
\begin{equation} \label{Appendix-CLF-HOCBF-case4-6}
	\mathbf{u}
	=-\frac{\Gamma_v}{1/\rho+\omega_2}\mathbf{\mathcal{L}_g}{\mathbf{\mathbf{\phi}}}^T_{r-1}.
\end{equation}
In this case $x\in \Omega_2$. \par
When $\mathbf{\mathcal{L}_g}{{\mathcal{{\mathcal{{\mathcal{\varphi}}}}}}}_{r-1}\neq 0$ , we have $\Gamma_v \neq 0$. Then, $\mu_1$, $\mu_2$ can be calculated as
\begin{equation} \label{Appendix-CLF-HOCBF-case4-7}
	\mu_1=\Delta^{-1}(\Gamma_v\mathbf{\mathcal{L}_g}{{\mathcal{{\mathcal{{\mathcal{\varphi}}}}}}}_{r-1}\mathbf{\mathcal{L}_g}{{\mathcal{{\mathcal{{\mathcal{\varphi}}}}}}}^T_{r-1}-\Gamma_b\mathbf{\mathcal{L}_g}{\mathbf{\mathbf{\phi}}}_{r-1}\mathbf{\mathcal{L}_g}{{\mathcal{{\mathcal{{\mathcal{\varphi}}}}}}}^T_{r-1}),
\end{equation}
\begin{equation} \label{Appendix-CLF-HOCBF-case4-8}
	\mu_2=\Delta^{-1}(\Gamma_v\mathbf{\mathcal{L}_g}{\mathbf{\mathbf{\phi}}}_{r-1}\mathbf{\mathcal{L}_g}{{\mathcal{{\mathcal{{\mathcal{\varphi}}}}}}}^T_{r-1}-\Gamma_b(1/\rho+\mathbf{\mathcal{L}_g}{{\mathcal{{\mathcal{{\mathcal{\varphi}}}}}}}_{r-1}\mathbf{\mathcal{L}_g}{{\mathcal{{\mathcal{{\mathcal{\varphi}}}}}}}^T_{r-1})).
\end{equation}
From \eqref{Appendix-CLF-HOCBF-1}, we obtain $\mathbf{u}=-\mu_1\mathbf{\mathcal{L}_g}{\mathbf{\mathbf{\phi}}}^T_{r-1}+\mu_2\mathbf{\mathcal{L}_g}{{\mathcal{{\mathcal{{\mathcal{\varphi}}}}}}}^T_{r-1}$, with $\mu_1$, $\mu_2$ are defined in \eqref{Appendix-CLF-HOCBF-case4-7}, \eqref{Appendix-CLF-HOCBF-case4-8}, respectively. In this case, $x\in \Omega_6$.
\section{Proof of Theorem \ref{Th-HOCBF-1}} \label{Appendix-B} %
 The Lagrangian associated to the QP \eqref{HOCBF-Control-1} is $\ell_{2}=\frac{1}{2}\| \mathbf{u}-\mathbf{\tau}_n(\mathbf{x})\|^2_2-\mu_c(\Gamma_b+\mathbf{\mathcal{L}_g}{{\mathcal{{\mathcal{{\mathcal{\varphi}}}}}}}_{r-1}\mathbf{u})$, where $\mu_c$ are the Lagrangian multipliers. The KKT conditions are:
\begin{equation} \label{Appendix-HOCBF-1}
	\begin{aligned}%
		\frac{\partial{\ell_{2}}}{\partial{\mathbf{u}}}=\mathbf{u}-\mathbf{\tau_n}-\mu_c\mathbf{\mathcal{L}_g}{\mathcal{\varphi}}^T_{r-1}=0,
	\end{aligned}
\end{equation}
\begin{equation} \label{Appendix-HOCBF-2}
\begin{aligned}%
\mu_c(\Gamma_b+\mathbf{\mathcal{L}_g}{\mathbf{\mathbf{\varphi}}}^T_{r-1}u)=0,
\end{aligned}
\end{equation}
\begin{equation} \label{Appendix-HOCBF-3}
\begin{aligned}%
\mu_c\ge0,
\end{aligned}
\end{equation}
\begin{equation} \label{HOCBF-Control-4}
\begin{aligned}
\mathcal{L}_f{\mathcal{\varphi}}_{r-1}(\mathbf{x})+\mathbf{\mathcal{L}_g}{\mathcal{\varphi}}_{r-1}(\mathbf{x})\mathbf{u}+\beta_r({\mathcal{{\mathcal{{\mathcal{\varphi}}}}}}_{r-1}(\mathbf{x}))\ge 0.
\end{aligned}
\end{equation}
According to \eqref{Appendix-HOCBF-1}, we have 
\begin{equation} \label{Appendix-HOCBF-4}
\begin{aligned}%
\mathbf{u}=\mathbf{\tau_n}+\mu_c\mathbf{\mathcal{L}_g}\mathcal{\varphi}^T_{r-1}.
\end{aligned}
\end{equation}
When $\mathbf{\mathcal{L}_g}\mathcal{\varphi}_{r-1}\mathbf{\tau_{n}}+\Gamma_b(\mathbf{x}) \ge 0$, the HOCBF constraints is inactive. In this case, $\mu_c=0$. Thus, we have $\mathbf{\tau}_{QP}=\tau_{n}$. When $\mathbf{\mathcal{L}_g}\mathcal{\varphi}_{r-1}=0$, from \eqref{Appendix-HOCBF-1} we have  $\mathbf{\tau}_{QP}=\tau_{n}$.\par
When $\mathbf{\mathcal{L}_g}\mathcal{\varphi}_{r-1}\tau_{n}+\Gamma_b < 0$, the HOCBF constraints is active. Suppose that
\begin{equation} \label{Appendix-HOCBF-5}
	\begin{aligned}%
		\Gamma_b+\mathbf{\mathcal{L}_g}{\mathbf{\mathbf{\varphi}}}^T_{r-1}\mathbf{u}=0.
	\end{aligned}
\end{equation}
Submitting \eqref{Appendix-HOCBF-4} into \eqref{Appendix-HOCBF-5} yields
\begin{equation} \label{Appendix-HOCBF-6}
	\begin{aligned}%
		\Gamma_b+\mathbf{\mathcal{L}_g}{\mathbf{\mathbf{\varphi}}}^T_{r-1}\mathbf{\tau}_n+\mu_c\|\mathbf{\mathcal{L}_g}{\mathbf{\mathbf{\varphi}}}^T_{r-1}\|^2_2=0.
	\end{aligned}
\end{equation}
Then, we can get 
\begin{equation} \label{Appendix-HOCBF-7}
	\begin{aligned}%
		\mu_c=-\frac{\Gamma_b+\mathbf{\mathcal{L}_g}{\mathbf{\mathbf{\varphi}}}^T_{r-1}\mathbf{\tau}_n}{\|\mathbf{\mathcal{L}_g}{\mathbf{\mathbf{\varphi}}}^T_{r-1}\|^2_2}.
	\end{aligned}
\end{equation}
Submitting \eqref{Appendix-HOCBF-7} into  \eqref{Appendix-HOCBF-4}, we have
\begin{equation} \label{Appendix-HOCBF-8}
	\begin{aligned}%
		\mathbf{u}=\mathbf{\tau}_n-\frac{\Gamma_b+\mathbf{\mathcal{L}_g}{\mathbf{\mathbf{\varphi}}}^T_{r-1}\mathbf{\tau}_n}{\|\mathbf{\mathcal{L}_g}{\mathbf{\mathbf{\varphi}}}^T_{r-1}\|^2_2}\mathbf{\mathcal{L}_g}{\mathbf{\mathbf{\varphi}}}^T_{r-1}.
	\end{aligned}
\end{equation}
Thus, we can obtain the solution in \eqref{HOCBF-sol} and \eqref{HOCBF-Control-2}.
\section{Proof of Theorem \ref{Th-TISSf-HOCBF-ForwardInvariant}} \label{Appendix-C} %
For a controller satisfying $\mathbf{u}\in \mathcal{K}_{TISSf}(\mathbf{x})$, we have
\begin{equation} \label{Appendix-TISSf-HOCBF-1}
\begin{aligned}%
\mathcal{L}_f\mathcal{\varphi}_{r-1}&+\mathbf{\mathcal{L}_{g1}}\mathcal{\varphi}_{r-1}\mathbf{u}+\mathbf{\mathcal{L}_{g2}}\mathcal{\varphi}_{r-1}\mathbf{d}
+\beta_r(\mathcal{\varphi}_{r-1})\\
\ge& \frac{\|\mathbf{\mathcal{L}_{g2}}{{\mathcal{{\mathcal{{\mathcal{\varphi}}}}}}}_{r-1}\|^2_2}{\epsilon(\mathcal{\varphi}_{r-1})}+\mathbf{\mathcal{L}_{g2}}{\mathcal{\varphi}}_{r-1}\mathbf{d}.
\end{aligned}
\end{equation}
It is noticed that the following inequation holds
\begin{equation} \label{Appendix-TISSf-HOCBF-2}
	\begin{aligned}%
		\mathbf{\mathcal{L}_{g2}}{{\mathcal{{\mathcal{{\mathcal{\varphi}}}}}}}_{r-1}\mathbf{d}&\ge-\|\mathbf{\mathcal{L}_{g2}}\mathcal{\varphi}_{r-1}\|\|\mathbf{d}\|_\infty \ge -\|\mathbf{\mathcal{L}_{g2}}\mathcal{\varphi}_{r-1}\|\gamma,
	\end{aligned}
\end{equation}
and $\epsilon(B(\mathbf{x}))>0$ $\forall \mathbf{x}\in \mathbf{R}^n$, $t\in[0,\infty]$.
Submitting \eqref{Appendix-TISSf-HOCBF-2} into \eqref{Appendix-TISSf-HOCBF-1}, we have
\begin{equation} \label{Appendix-TISSf-HOCBF-3}
\begin{aligned}%
\mathcal{L}_f\mathcal{\varphi}_{r-1}&+\mathbf{\mathcal{L}_{g1}}\mathcal{\varphi}_{r-1}\mathbf{u}+\mathbf{\mathcal{L}_{g2}}\mathcal{\varphi}_{r-1}\mathbf{d}
+\beta_r({\mathcal{{\mathcal{{\mathcal{\varphi}}}}}}_{r-1})\\
\ge& \frac{\|\mathbf{\mathcal{L}_{g2}}{{\mathcal{{\mathcal{{\mathcal{\varphi}}}}}}}_{r-1}\|^2_2}{\epsilon(\mathcal{\varphi}_{r-1})}-\|\mathbf{\mathcal{L}_{g2}}\mathcal{\varphi}_{r-1}\|\gamma \ge -\frac{\epsilon(\mathcal{\varphi}_{r-1})\gamma^2}{4}.
\end{aligned}
\end{equation}
Denote $B^{\gamma}(\mathbf{x})=B(\mathbf{x})+\varrho(B(\mathbf{x}), \gamma)$, ${\mathcal{\varphi}}^{\gamma}_{r}(\mathbf{x})={\mathcal{\varphi}}(\mathbf{x})+\varrho(\varphi(\mathbf{x}), \gamma)$. The time derivative of ${\mathcal{\varphi}}$ can be obtained as 
\begin{equation} \label{Appendix-TISSf-HOCBF-4}
	\begin{aligned}%
		{\mathcal{\dot{\varphi}}}^{\gamma}_{r-1}=\big(1+\frac{\partial {\varrho}}{\partial{\mathcal{\varphi}_{r-1}}}(\mathcal{\varphi}_{r-1}, \gamma)\big){\mathcal{\dot{\varphi}}}_{r-1}.
	\end{aligned}
\end{equation}
As $\epsilon $ satisfies \eqref{TISSf-HOCBF-Control} and $\varrho$ defined in  \eqref{TISSf-HOCBF-varrho}, we have 
\begin{equation} \label{Appendix-TISSf-HOCBF-5}
	\begin{aligned}%
		1+\frac{\partial {\varrho}}{\partial{\mathcal{\varphi}_{r-1}}}(\mathcal{\varphi}_{r-1}, \gamma)>0.
	\end{aligned}
\end{equation}
Substituting \eqref{Appendix-TISSf-HOCBF-3} into \eqref{Appendix-TISSf-HOCBF-4} yields:
\begin{equation} \label{Appendix-TISSf-HOCBF-6}
\begin{aligned}
{\mathcal{\dot{\varphi}}}^{\gamma}_{r}\ge&\big(1+\frac{\partial {\varrho}}{\partial{\mathcal{\varphi}_{r-1}}}(\mathcal{\varphi}_{r-1}, \gamma)\big)\big(-\beta_r({\mathcal{{\mathcal{{\mathcal{\varphi}}}}}}_{r-1})-\frac{\epsilon(\mathcal{\varphi}_{r-1})\gamma^2}{4}\big).
\end{aligned}
\end{equation}
Considering the fact that:
\begin{equation} \label{Appendix-TISSf-HOCBF-7}
\begin{aligned}
-\beta_r&({\mathcal{{\mathcal{{\mathcal{\varphi}}}}}}_{r-1})-\frac{\epsilon(\mathcal{\varphi}_{r-1})\gamma^2}{4}\\
&=-\big((\mathcal{\varphi}_{r-1})+\beta^{-1}_r\big(-\frac{\epsilon(\mathcal{\varphi}_{r-1})\gamma^2}{4}\big)\big)\\
&=-{\mathcal{\varphi}}^{\gamma}_{r-1}.
\end{aligned}
\end{equation}
Submitting \eqref{Appendix-TISSf-HOCBF-7} into \eqref{Appendix-TISSf-HOCBF-6}, we can obtain:
\begin{equation} \label{Appendix-TISSf-HOCBF-8}
\begin{aligned}
{\mathcal{\dot{\varphi}}}^{\gamma}_{r}\ge
&-\Gamma{\mathcal{\varphi}}^{\gamma}_{r-1}, 
\end{aligned}
\end{equation}
where $\Gamma=\big(1+\frac{\partial {\varrho}}{\partial{\mathcal{\varphi}_{r-1}}}(\mathcal{\varphi}_{r-1}, \gamma)\big)$. Since $\Gamma{\mathcal{\varphi}}^{\gamma}_{r-1} \in \Gamma(\mathcal{\varphi}^{\gamma}_{r-1})$ is a class $\mathcal{K}_{\infty}$ function. Thus, we have system \eqref{System-Fun-d} is ISSf on $\mathcal{C}$, and the set $\mathcal{C}_{\gamma}$ robustly forward invariant.
%
\section*{References}
\bibliographystyle{IEEEtran}        
\bibliography{autosam}  
\vspace{-1cm}
\end{document}